# Wormholes, Void Bubbles and Vacuum Energy Suppression

Enrico Rodrigo
*Department of Physics and Astronomy, University of California at Irvine,
Irvine, CA 92697-4575, USA*
and
*Department of General Studies, Charles Drew University
Los Angeles, CA 90059, USA*

The gargantuan discrepancy between the value of the observed cosmological constant and that expected from the zero-point energy of known matter fields can be eliminated by supposing that on macroscopic scales the overwhelming majority of any volume of spacetime is literal nothingness. This nothingness or void results from the proliferative nucleation of tiny void bubbles (a.k.a. "bubbles of nothing" or "semi wormholes") that expand until their surfaces, presumed to be 2-branes, collide. This process results in a dense packing of void bubbles of various sizes that leaves only the vanishing interstitial regions between bubbles for spacetime to occupy. This vast reduction in the amount of actual space contained within any apparent volume, reduces correspondingly the effective zero-point energy density. Unlike previous wormhole-based attempts at vacuum energy suppression, the current approach is entirely Lorentzian and results in a nonzero value for the cosmological constant.

PACS Number(s): 04.20.Gz, 04.60.-m, 04.60.Ds

## 1. Introduction

The zero-point energy density of known matter fields can be up to fifty-five orders of magnitude larger than the energy density corresponding to the observed value of the cosmological constant. This is the cosmological-constant problem, although it is now sometimes referred to as the "old cosmological constant problem" (see [1-6] for reviews). Over the years numerous mechanisms have been proposed for solving it. None have become widely accepted (see [33-38] for a few recent approaches). While most recent activity seems to be inspired by KKLT [15], I would like to focus instead on a new variation of two approaches that by now may be considered "golden oldies": wormholes and bubble nucleation.

Coleman's wormhole approach appeared in 1988 [7] (see also [10] for a play-by-play description of the key derivation). He considered the formal solution to the Wheeler-DeWitt equation expressed as the usual Euclidean path integral of *exp*(Action) over the set of all possible geometries. This path integral is, of course, only tractable in a minisuperspace or other approximation that reduces the "set of all possible geometries" to a set spanned by a finite number of parameters. Coleman chose in effect the set of geometries consisting of distinct macroscopic 4-spheres connected to each other by arbitrary numbers of wormholes. He argued that the corresponding Euclidean path integral is dominated by contributions from geometries in which the diameters of the connecting wormholes were of Planckian scale. He showed, moreover, that this integral – the amplitude for various geometries – peaks when the effective cosmological constant is precisely zero. Irrespective of our current belief that "precisely zero" is the wrong answer and irrespective also of the questionable relevance of Euclidean path integrals to our manifestly Lorentzian universe, Coleman's approach was shown to be problematic. It turns out there are significant contributions to the path integral from geometries featuring wormholes of all diameters [9, 8]. The macroscopically porous Swiss-cheese universe implied by this "wormhole catastrophe" conflicts with our most casual observations.

Three years earlier Abbot had exploited another of Coleman's ideas – that of false vacuum decay [12] – to propose the first bubble nucleation mechanism for addressing the problem [11]. Abbot realized that if the potential function characterizing a scalar field was periodic with descending local minima, the potential would permit useful tunneling between the minima. This sort of potential function also ensures that tunneling occurs preferentially in the direction that lowers the field's vacuum energy and thus the value of the cosmological constant. Each jump is an instance of false vacuum decay, which manifests itself as a nucleation and rapid growth of a bubble whose interior features a lower cosmological constant. Two years after Abbot (and a year before Coleman's wormholes) Brown and Teitelboim proposed an alternative bubble nucleation scheme. They linked the value of the cosmological constant to the energy density of a four-form (a totally antisymmetric tensor) field. When a bubble nucleates, this energy density is reduced within it for a simple reason. Energy must be subtracted from the bubble's interior in order to form its surrounding membrane. The bubble rapidly expands and creates a region of reduced four-form energy density in which new bubbles form enclosing regions of even lower energy density. This process continues until the energy density, i.e. the cosmological constant, is too low to permit further nucleation. Brown and Teitelboim recognized that the creation and growth of bubbles at the speed of light occurs too slowly to be consistent with conventional cosmology. They understood, moreover, that their process requires an extraordinarily fine discretization of matter charges on the bubble membranes, and that it renders the universe cold and lifeless. Some of these issues were addressed several years ago [13, 14], when interest in this mechanism was renewed by the natural occurrence within string theory of four-form fields suitably coupled to branes.

The purpose of this note is to show that these two ideas – wormholes and bubble nucleation – taken together suggest another approach free of the aforementioned drawbacks.

## 2. The idea summarized

Consider a Lorentzian de Sitter space with cosmological constant $\Lambda$ and an associated energy density $\rho_{vac}$ that are due solely to the zero point energy of quantum fields. In the absence of a compelling reason for believing that a quantum theory of gravity will be articulated as a quantum field theory, I will not presume $\rho_{vac}$ to have a Planck-scale contribution. Rather, I will take $\rho_{vac}$ to consist entirely of zero-point contributions from known matter fields and to therefore be of the electroweak scale of about $(100 \text{ GeV})^4$. Imagine that an unspecified global event increases from zero to some finite value the probability per unit volume per unit time of bubble nucleation. Suppose further that this nucleation is of a special type. The nascent bubble, instead of enclosing a new region of reduced $\Lambda$, encloses nothing – a void. In other words, the bubbles are semi wormholes with 2-branes at their throats. Hence, these 2-branes are boundaries of spacetime. Because they are Dirichlet, the values of all fields propagating within the spacetime are determined at these boundaries. This preserves the predictability of physics within the spacetime's natural Cauchy horizons [16]. Here, unlike the usual form of bubble nucleation, the creation of a bubble is tantamount to a change in the effective topology of spacetime – the sudden appearance of a growing hole. Although I shall only consider (3+1)-dimensional wormholes with 2-branes at their throats, the analysis can be extended in the context of brane world models to include higher-dimensional wormholes [32].

The throats of these semi-wormholes, which I shall henceforth refer to as "void bubbles", will continue to expand until they collide. Were these ordinary wormhole throats, they would likely coalesce. As throats coincident with suitably stiff branes, the colliding void bubbles halt their expansions and become densely packed throughout the whole of space. This reduces the volume of space to that of the interstitial region between the void bubbles. Smaller bubbles will nucleate within these regions, further reducing the interstitial volume. As the process continues, progressively smaller void bubbles will form within progressively reduced interstitial regions. Were this to continue indefinitely, the entire volume of space would be consumed. Given that this could not have occurred, the nucleation process must at some point have terminated.

To quantify the terminal state of the nucleation process, define an "apparent volume" of a region of spacelike hypersurface as follows: the volume calculated by an observer who, ignorant of the presence of void bubbles, assumes the region to be wholly occupied by simply connected normal space with the observed background metric. Hence, in Minkowski space a cube with sides of length $d$ containing a void bubble of radius $d/2$ has an apparent volume of $d^3$, although it only contains a volume of normal space equal to
$d^3-(4\pi/3)(d/2)^3$.

A measure of the degree to which the nucleation process succeeded in consuming normal space is the fraction of such space that remains within any apparent volume of a spacelike hypersurface. For every Planck volume $L_P^3$ of normal space that survives consumption by bubbles, there exists a much larger volume of normal space that does not. We can, moreover, assume that the great majority of the space consumed within a suitably sized apparent volume was swallowed by a single bubble of radius $s$ – one of the earliest bubbles to form after the global event that triggered mass nucleation. Recall that later bubbles merely consumed the interstitial regions left between early bubbles, each of whose expansion was halted by that of their early neighbors. The ratio, then, of the volume of remaining normal space to that consumed is roughly

$$\frac{L_P^3}{s^3}. \tag{1}$$

As the void, which occupies nearly all of the apparent volume of interest, contains no zero-point energy, the vacuum energy of the apparent volume is due solely to its remaining normal space. This drastically reduces the vacuum energy per apparent volume. In terms of the electroweak zero-point energy density $\rho_{vac}$ and the observed density $\rho_{obs}$, we have

$$\frac{\rho_{obs}}{\rho_{vac}} = \frac{\frac{\rho_{vac} L_P^3}{s^3}}{\rho_{vac}} = \frac{L_P^3}{s^3} = 10^{-55} \tag{2}$$

where the apparent volume of the region under consideration is assumed to be $s^3$ -- about that of the largest void bubble that it contains. These bubbles, then, have an average radius $s$ of about $10^{-15}$ cm.

One might think that this radius should be somewhat reduced by the Casimir effect, whose likely consequence (besides promoting bubble nucleation) is to reduce $\rho_{vac}$ in the numerator of eq. (2) through the exclusion of modes whose wavelengths are too long to fit within the remaining interstitial volume. However, the seemingly separate pockets of interstitial volume between the dominant void bubbles are *not* isolated from each other. It is not at all clear that long-wavelength vacuum modes are prevented from contributing to $\rho_{vac}$. This view is reinforced, when we regard the remaining interstitial volume as a sort of irregular lattice. There is no obvious reason why such a lattice, despite being characterized by detailed features at the Planck-length scale, could not support modes of the electro-weak length scale and longer. It might seem prudent, therefore, to consider the worst case -- in the sense that it results in the largest radius for the dominant bubbles – by ignoring the Casimir effect.

However, a non-existent Casimir effect may only be considered a worst case in this scenario, if the Casimir energy is necessarily negative, i.e. if it is only able to reduce $\rho_{vac}$ in the numerator of eq. (2). A negative-definite Casimir energy is not assured, given that the Casimir energy is known to be *positive* in certain situations, most notably that of an isolated perfectly conducting sphere (see for example [41] and the references therein). Because our void bubbles are spherical, it might seem that the Casimir energy for such a sphere would be an apt choice to describe the Casimir force between the bubbles. It is not, however, because the void bubbles are not isolated. A superior choice would be the Casimir energy between two conducting (Dirichlet) spheres. This was calculated recently for the case of a scalar field [42]. When the spheres are in close proximity – when their separation is roughly less than their radii – their Casimir energy is negative. As they separate (and thus become isolated), their Casimir energy becomes positive. There currently exists no method for calculating the Casimir effect for arbitrary geometries, and research in this area remains active. To the degree, however, that contiguous void bubbles resemble conducting spheres in closes proximity, it appears that their interstitial Casimir energy is negative. A negative Casimir energy only acts (in the numerator of eq.(2)) to reduce the size of the dominant void bubbles. It is therefore likely that ignoring the Casimir effect altogether, will better strain our thesis by yielding an upper limit (a worst case) on the size of the dominant void bubbles. Accordingly, I will ignore the Casimir effect in what follows.

The picture painted above is one of spacetime consisting of an irregular lattice formed by a dense packing of submicroscopic void bubbles between which are ensconced the minute slivers of normal space that constitute the lattice and in which matter fields reside (Figure 1). In short, the chief constituent of space, by an overwhelming majority, is literal nothingness. Spacetime is hollow. This suggests that the solution to the cosmological constant problem is the realization that the quantity of space assumed to contribute to the vacuum energy within a given volume does not in fact exist – it merely appears to. Just as matter despite appearances is mostly empty space, empty space is mostly void.

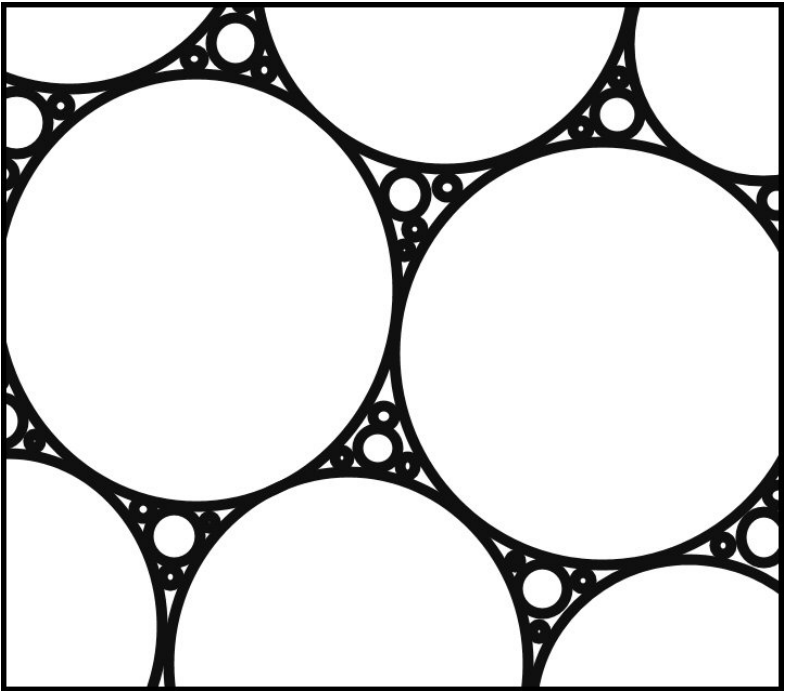

**Figure 1. Void bubbles in spacetime.** Early forming void bubbles are the largest and replace the bulk of spacetime with void. In this sense they are dominant. Progressively smaller bubbles consume nearly all of the remaining normal space within the interstices of the dominant bubbles. What remains of normal space exists as unconsumed Planck-scale bits within the interstices.

**3. Two immediate concerns**

Is a dominant void bubble size of $10^{-15}$ cm consistent with the current accelerator resolution of $10^{-17}$ cm? Recent developments in high-precision lattice models suggest that it is. These models, which typically involve lattice spacing of about $10^{-14}$ cm, are in excellent agreement with experiment [17, 18]. This agreement, moreover, is insensitive to variations in lattice spacings in this range. This insensitivity is usually interpreted as an indication of the model's fidelity to continuum physics. If the assumption of a continuum is abandoned, however, the insensitivity may be taken instead to mean that the scale of the effective discretization of spacetime is near that of the model.

Because void bubbles are semi-wormholes, each with a 2-brane at its throat, they might be supposed to necessarily possess a positive mass. Classically, at least, this is not the case. A wormhole's ADM mass [29] bears no direct dependence on the value of the stress energy tensor in the vicinity of its throat [30]. To be sure, the assumed absence of an event horizon (the chief criterion of wormhole traversability) requires the matter near the throat to violate the weak energy condition [27]. But this would only require the tension of the brane to be negative, assuming that its equation of state remains $p = -\rho$. It would not prevent the Riemann tensor from vanishing beyond the immediate vicinity of the throat. It would not, in other words, prevent the void bubble from being massless in the ADM sense (or, to be more precise, the generalization of the ADM concept to an asymptotically de Sitter space). The extent to which quantum considerations or internal degrees of freedom would allow the bubbles to absorb energy and thereby acquire a slight mass is unknown. However, it is clear that the issue is of immediate relevance to cosmology – and in particular the identification of a possible constituent of dark matter.

## 4. The model

Our task now is to develop a simple model that permits void bubble nucleation resulting in a dominant bubble radius of $10^{-15}$ cm. We can accomplish this by resorting to a minisuperspace treatment. Its central feature is a function of the bubble's radius that acts as an effective potential driving the bubble's dynamics. When the bubble is much larger than a Planck length, this potential function can be determined by the Darmois-Israel thin shell formalism [19, 20]. For Planck-scale bubbles this formalism will not apply due to the importance of higher order terms in the gravitational Lagrangian -- terms whose existence are inconsistent with the formalism's fundamental assumptions. The potential function can, nevertheless, be extended into the Planckian regime through the use of *ad hoc* assumptions inspired by Wheeler's intuition about the nature of spacetime foam [21]. With the potential function so determined, the system can be canonically quantized in a manner that circumvents the nonlocal operators that characterize the standard Wheeler-DeWitt approach. The resulting Schrödinger equation can then be solved approximately to determine parameter values corresponding to the nucleation of dominant bubbles of the aforementioned size.

We begin by recalling that a void bubble is a semi-wormhole. That is to say it is the spacetime in the immediate vicinity of one side of a wormhole throat together with the throat itself. The other half of the wormhole -- that on the other side of its throat (which presumably exists in another universe or a distant region of our universe) -- is to be discarded after we work out the wormhole's dynamics using the thin shell method [16]. In other words, a void bubble adheres to the same equation of motion as the corresponding full wormhole. We can construct our thin shell wormhole by surgically connecting two solutions to the Einstein equations – the first solution pertains to our universe, the second to the soon-to-be-discarded other universe. Choosing both solutions to be identical creates a wormhole symmetrical about it throat. A simple choice for a massless (in the ADM sense), uncharged wormhole in an asymptotically de Sitter spacetime is the solution

$$ds^2 = -\left(1-\frac{\Lambda}{3}r^2\right)dt^2 + \left(1-\frac{\Lambda}{3}r^2\right)^{-1}dr^2 + r^2 d\Omega. \tag{3}$$

To create the wormhole remove from each of two identical manifolds (on which this solution holds) a spherical hole of radius $a$ centered about the origin. Glue the manifolds together at $r=a$. The thin shell formalism together with the First Law of Thermodynamics and the equation of state $p=-\rho$ yields the Friedman-like equation [16]

$$\left(\frac{2\pi G \rho_0}{c^4}\right)^2 = \frac{1}{a^2} - \frac{\Lambda}{3} + \left(\frac{d\ln(a)}{cd\tau}\right)^2, \tag{4}$$

where $\rho_0$ is the initial density at the wormhole's throat, $a$ is the throat's radius, $\tau$ is proper time measured by an observer at the throat, $G$ is the gravitation constant, and $c$ the speed of light. The expected presence of higher order terms in perhaps even the classical Lagrangian for gravity means that eq. (4) cannot hold for arbitrarily small wormhole radii. The higher order curvature terms would invalidate the Darmois-Israel thin-shell formalism (by introducing quadratic and higher powers of the delta-function in the stress energy tensor that would not integrate out). One means of extending (4) into the regime of small radii is simply to guess. If something like spacetime foam exists, we might suppose that this is only possible in an extension of eq. (4) if this foam is realized as a sort bound state. Accordingly, we replace eq. (4) with

$$\left(\frac{2\pi G \rho_0}{c^4}\right)^2 = U(a) + \left(\frac{d \ln(a)}{cd\tau}\right)^2 \tag{5}$$

where

$$\begin{aligned} U(a) &= \infty & \text{for } a \leq a_0 \\ U(a) &= U_0 & \text{for } a_0 < a < a_1 \\ U(a) &= a^{-2} - \Lambda/3 & \text{for } a > a_1 \end{aligned} \tag{6}$$

and $U_0$ is a negative constant (Figure 2). The lengths $a_0$ and $a_1$ are respectively the inner and outer boundaries of a square well potential (Figure 2). We take $a_0$ to be on the order of a Planck length and $a_1$ to be one or two orders of magnitude larger.

The essential features of the guessed portion of the potential given by eqs. (6) -- the existence of a flat minimum and a divergence at $a = a_0$ -- are not entirely without justification. The effective potential determining the dynamics of a quantum cosmology governed by $R + R^2$ gravity (mathematically similar to the analogous quantum wormhole) shows a similar minimum that is insensitive to the value of the radius parameter, when the parameter is small [39]. The potential also features a divergence corresponding to the "bounce" that occurs when the radius reaches its minimum value [39]. This divergence likely indicates in both cases the impossibility of genuine topology change.

Our assumptions about brane-stress are implicit in our choice for the equation of state (for the classical and semiclassical regimes) and in the value of the constant $U_0$ (for the quantum regime). The model will retain its qualitative features as long as the sign of brane tension is the same as that of brane's energy density, and as long as $U_0$ is negative. Other choices for the brane tension would require other choices for the equation of state parameter $w$ in $p = w\rho$ and other choices for $U_0$. The model, then, is consistent with non-zero brane stress and incorporates it precisely as in the standard treatment of thin-shelled wormholes [See [43] for a textbook account.].

The $\Lambda$ that appears in the potential $U(a)$ of eq. (6) is the cosmological constant that corresponds to the electroweak vacuum, i.e. $\Lambda = \Lambda_{vac.} = (8\pi G/c^4)\rho_{vac}$. Hence the "energy spectrum" and barrier penetration probabilities that we shall determine below, will all depend upon $\Lambda_{vac}$. This $\Lambda$ will moreover determine the rate of void bubble nucleation per unit volume and thus the size of the dominant void bubbles that result when nucleation occurs en masse. It is only *after* such mass nucleation has occurred and the universe is thereby densely packed with void bubbles of dominant radius $s$, that the value of the observed cosmological constant is reduced to match the reduced value $\rho_{obs}$ of the energy density. This occurs as described in eq.(2). In short, $\Lambda_{obs} = (8\pi G/c^4) \rho_{obs} = (8\pi G/c^4)(L_P/s)^3 \rho_{vac}$, where, again, the estimated value for $s$ depends on the energy spectrum determined below and thus ultimately on $\Lambda_{vac.}$.

With the change of variable, $q = a_0 ln(a/a_0)$, eq. (5) assumes the form of the first integral of the equation of motion of a particle under the influence of a potential: a constant set to the sum of a kinetic and a potential term. Although eq. (5) fully describes the classical dynamics of the system, it is unsuitable for canonical quantization in its current form. Were it multiplied by a suitably dimensioned constant, it would become an energy balance equation of a new *classically indistinguishable* system. As such it would lend itself to the traditional quantization procedure. Such a multiplication would, however, introduce another length scale into the problem. If this length is chosen properly, one might suppose that quantization based on the inferred Hamiltonian of the new system could serve as a shortcut avoiding the relatively laborious standard approaches [see for example 22, 23] that begin with the cumbersome Hamiltonian of the gravitational field. This cheap quantization for the purpose of easily extracting the qualitative features of the system is not dissimilar to an alternative quantization of the Friedmann-Robertson-Walker cosmology that circumvents the usual Wheeler-DeWitt approach [24, 25, 32]. [See also [26] for a similar approach applied to wormholes.]

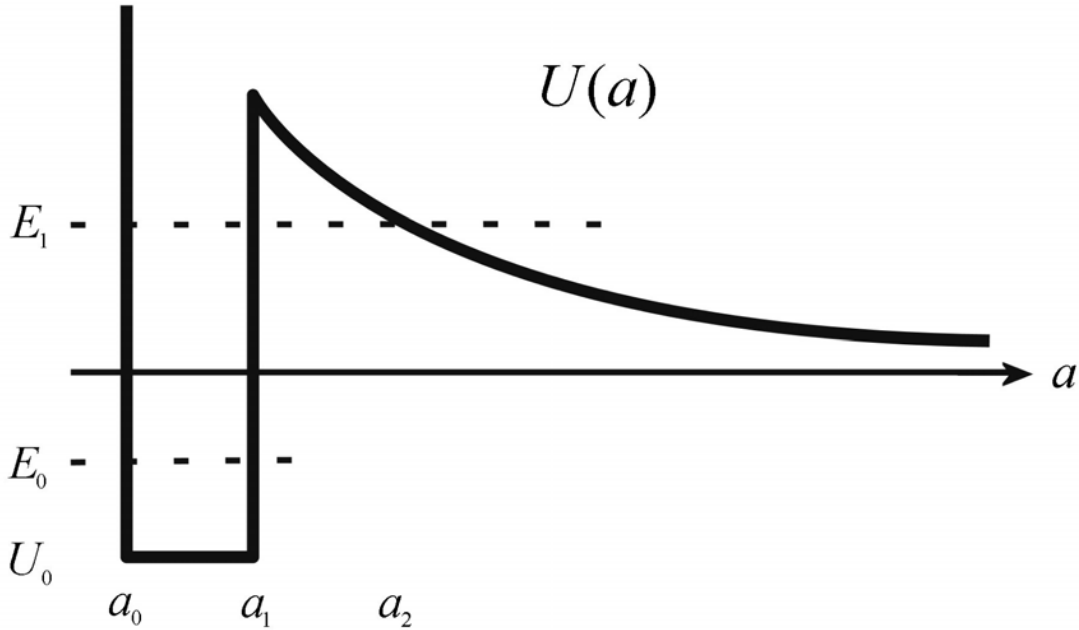

**Figure 2. Void bubble potential function.** For bubble radius $a > a_1$, thin shell formalism assumed valid. For $a < a_1$, thin shell formalism inapplicable due to the expected presence of higher-order terms in the Lagrangian of gravity. Effect of higher-order terms is guessed to be roughly similar to the square well shown. Form forbids bubble radii below $a_0$ thus preventing true topology change. Bubbles with WKB ground state energy $E_0$, may be regarded as existing within Wheeler's spacetime foam. Bubbles elevated to the first excited WKB state $E_1$ tunnel through the potential barrier and are driven to expand by the declining potential.

Choosing a length scale already built into the model, we multiply eq.(5) by $a_0^3 c^4/G$ to obtain

$$E = \tfrac{1}{2} m \dot{q}^2 + V(q) \qquad (7)$$

where

$$E \equiv \frac{4\pi^2 G a_0^3}{c^4} \rho_0^2 \quad m \equiv \frac{2 a_0 c^2}{G} \quad V(q) \equiv \frac{a_0^3 c^4}{G} U(a(q)) \quad (8)$$

and

$$q \equiv a_0 \ln(a/a_0). \qquad (9)$$

The usual quantization obtains the Schrödinger equation to which the application of the WKB approximation results in the Bohr-Sommerfeld quantization rule

$$\int_{q_0}^{q_1} \sqrt{2m(E-V_0)}\, dq = \left(n+\tfrac{1}{2}\right)\frac{h}{2} \qquad (10)$$

yielding

$$E_n = V_0 + \frac{\left(n+\tfrac{1}{2}\right)^2 h^2 G}{16 c^2 a_0^3 \ln^2(a_1/a_0)} \qquad n = 0, 1, 2,\ldots \qquad (11)$$

where $q_0 \equiv q(a_0) = 0$ and $q_1 \equiv q(a_1) = a_0 \ln(a_1/a_0)$. Identifying bound states as proto bubbles oscillating within the spacetime foam, the penetration factor $P$ becomes the probability per encounter with the potential barrier of a proto bubble bursting through to begin its explosive expansion:

$$P_n = \exp\!\left(-\frac{2}{\hbar}\int_{q_1}^{q_2}\sqrt{2m(V(q)-E_n)}\, dq\right). \qquad (12)$$

where $V(q_1) = V(q_2) = E_n$ and

$$\begin{aligned}
V(q) &= \infty & &\text{for } q \le q_0 \\
V(q) &= V_0 & &\text{for } q_0 < q < q_1 \\
V(q) &= \frac{a_0 c^4}{G} e^{-2q/a_0} - \frac{a_0^3 c^4}{G}\Lambda & &\text{for } q \ge q_1.
\end{aligned} \qquad (13)$$

Spacetime foam must of course be stable. This is tantamount to requiring $P_0=0$. If bubble nucleation is to occur, the penetration factor must exceed zero for an excited state. Choosing $P_1 > 0$, we see that these requirements on $P_0$ and $P_1$ constrain the depth $V_0$ of the potential well according to

$$\frac{-a_0^3 c^4 \Lambda}{G} - \frac{9 h^2 G}{64 c^2 a_0^3 \ln^2(a_1/a_0)} \le V_0 < \frac{-a_0^3 c^4 \Lambda}{G} - \frac{h^2 G}{64 c^2 a_0^3 \ln^2(a_1/a_0)}. \qquad (14)$$

If the system is in a superposition, $|\psi\rangle = \alpha|\psi_0\rangle + \sqrt{1-\alpha^2}\,|\psi_1\rangle$, of the two lowest eigenstates, the effective penetration factor $P$ becomes

$$P = \alpha^2 P_1 + (1-\alpha^2) P_0 = \alpha^2 P_1 \qquad 0 < \alpha < 1 \qquad (15)$$

where by (12) and (13)

$$P_1 = \exp\!\left\{\frac{-4 c^3 a_0^2}{\hbar G}\left[\sqrt{(a_0/a_1)^2 + b_1} - \sqrt{-b_1}\,\tan^{-1}\!\left(\sqrt{\frac{(a_0/a_1)^2 + b_1}{-b_1}}\right)\right]\right\} \qquad (16)$$

and

$$b_1 \equiv -a_0^2 \Lambda - \frac{G}{a_0 c^4} E_1. \tag{17}$$

This penetration factor will affect the dominant bubble radius. To see this, suppose that in some Lorentz frame we divide space into numerous submicroscopic cells each of which has the possibility of giving birth to a void bubble. An event occurs that induces many of these cell to birth an expanding bubble. The bubbles expand until their surfaces collide with those of their neighbors. Let $s$ be the average radius of the halted bubbles, $v$ the average radial speed of the surface of the expanding bubble, and $t_c$ the average time between a bubble's birth and the cessation of its expansion. Then

$$\frac{s}{v} \approx \frac{(\tilde{V}/N)^{1/3}}{v} = t_c, \tag{18}$$

where $\tilde{V}$ is an arbitrary volume, and $N$ is the number of bubbles born within this volume during the interval $t_c$. Let $\tilde{p}$ be the probability per unit time of a particular cell birthing a bubble and $\tilde{\rho}$ be the density of cells. Then $N = \tilde{p}\tilde{\rho}\tilde{V}t_c$. Inserting this in (18), solving for $t_c$, and eliminating it yields,

$$s = \left(\frac{v}{\tilde{p}\tilde{\rho}}\right)^{1/4} = \left(\frac{v\tau}{P\tilde{\rho}}\right)^{1/4}, \tag{19}$$

where $\tau$ is the period of oscillation of the void bubble in its bound state, which links the probability $\tilde{p}$ and the penetration factor $P$ by $\tilde{p} = P/\tau$. According to eq. (9)

$$\dot{q} = \frac{a_0}{a}\dot{a} \quad \text{or} \quad v_q = \frac{a_0}{a}v \tag{20}$$

The value of $a$ to use in the transformation (20) is the average value of $a$ during the expansion of the bubble: $a \approx s/2$. Using this choice with (20) and (19), we have

$$s = \left(\frac{v_q \tau a_1^3}{2 P a_0}\right)^{1/3} \tag{21}$$

where the spatial density $\tilde{\rho}$ of proto bubble cells is assumed to have its maximum value, i.e. $\tilde{\rho} = a_1^{-3}$. The dominant bubble radius, then, varies inversely with the cube root of the penetration factor. Note also that equations (18) and (19) hold even in the event of spatial expansion induced by a positive cosmological constant.

Because the potential $V(q)$ falls off rapidly, we may take $v_q$ to be its asymptotic value. This is easily obtained by finding the quantity of kinetic energy that remains after the bound "particle" with energy $E_1$ tunnels through the barrier. Given $E_1$ and the width $(q(a_1) - q(a_0))$ of the square well, we may eliminate $\tau$ in (21) and recall (15) to obtain

$$s = a_1 \left[\frac{\ln(a_1/a_0)}{2\alpha^2 P_1}\right]^{1/3} \left[1 + \frac{64 a_0^3 c^2 \ln^2(a_1/a_0)}{9 G h^2}\left(V_0 + \frac{a_0^3 c^4}{G}\Lambda\right)\right]^{1/6} \quad (22)$$

or, equivalently,

$$\alpha^2 = \frac{a_1^3 \ln(a_1/a_0)}{2 s^3 P_1}\left[1 + \frac{64 a_0^3 c^2 \ln^2(a_1/a_0)}{9 G h^2}\left(V_0 + \frac{a_0^3 c^4}{G}\Lambda\right)\right]^{1/2}. \quad (23)$$

We may now chose the parameters $a_1$, $a_0$, and $V_0$, set the dominant bubble radius $s$ to the value determined by eq. (2), use the value of $\Lambda$ corresponding to the electroweak vacuum and thus obtain through (23) the probability $\alpha^2$ required that a proto bubble be in its first excited state. Let $a_0 = L_P$, $a_1 = 10 a_0$, $s = 10^{-15}$ cm, and $\Lambda = 0.02$ cm$^{-2}$ (the value of $\Lambda$ corresponding to the electroweak $\rho_{vac} = 10^{46}$ ergs/cm$^3$). The constraint (14) requires $V_0$ to be of the form

$$V_0 = -\frac{a_0^3 c^4}{G}\Lambda - \beta \frac{h^2 G}{64 c^2 a_0^3 \ln^2(a_1/a_0)}, \quad (24)$$

where $9 \leq \beta < 1$. The observed stability of spacetime suggests a deep well for our foam. However, setting $\beta$ to precisely 9 (in order to obtain the deepest well) results in an infinite outer turning point $a(q_2)$ for the integrand in (12), which corresponds to the nucleation of infinitely large bubbles. Choose instead $\beta = 9 - \varepsilon^2$, for which the outer turning point is given by

$$a(q_2) = \frac{a_0}{\sqrt{-b_1}} = \frac{8 a_0^3 c^3 \ln(a_1/a_0)}{h G \varepsilon}. \quad (25)$$

Choosing $\varepsilon = 0.1$ to join our list of parameter values, we find the scale of freshly nucleated bubbles to be under 30 Planck lengths. And by eqs. (24) and (23) the probability $\alpha^2$ of the first excited state must be

$$\alpha^2 = 2.0 \times 10^{-52}. \quad (26)$$

Assuming that the proto bubble cells are distributed in a manner not entirely dissimilar to Boltzmann statistics, the temperature $T_N$ corresponding to this probability is given by

$$e^{-(E_1-E_0)/k T_N} = \frac{\alpha^2}{1-\alpha^2}$$

or

$$T_N = \frac{-h^2 G}{8 c^2 a_0^3 \ln^2(a_1/a_2) k \ln(\alpha^2)} \sim 10^{30} \text{ degrees K} \quad (27)$$

confirming that, despite the low value of $\alpha^2$, mass void bubble nucleation must have been triggered immediately after the big bang.

This seems to suggest a sequence of events. Inflation -- starting a bit earlier and at higher initial temperature than in the usual models – not only converts the vacuum energy of the inflaton into the inflationary expansion of the universe, but also raises the universe (viewed as a huge collection of potential bubble nucleation sources) from the vacuum state $|\psi_0\rangle$ to the aforementioned superposition $|\psi\rangle = \alpha |\psi_0\rangle + \sqrt{1-\alpha^2} |\psi_1\rangle$. Achieving this state, which the universe reaches at the end of its inflationary period, starts the bubble nucleation described above. This results in dominant void bubbles of about $10^{-15}$ cm and in smaller bubbles in the interstices between the dominant bubbles. Each interstitial bubble will attempt to expand, driven by the gradient $U'(a)$ until its expansion is halted by the combined pressure of its neighboring bubbles. This prevents newer bubbles from exceeding the size of the dominant bubbles. As the universe expands, bubbles nucleate into the newly created space. By conjecture, nucleation continues until one Planck volume of normal space per dominant void bubble remains. Dominant bubbles attempting to expand into the new space are thwarted by the superior expansive pressure (due to $U'(a)$ increasing with decreasing bubble radius) of newly created or previously existing bubbles in the interstices. Thus dominant void bubbles, like elementary particles and other effectively bound states, retain their sizes even as the post inflationary universe gradually expands by 27 orders of magnitude to its current volume.

## 5. Discussion

If spacetime is in fact congested with void bubbles, an immediate consequence, it would seem, would be the dispersion of radiation whose wavelengths are shorter than the maximum bubble size. Bubbles no larger than $10^{-15}$ cm would, presumably, cause an angular dispersion of radiation and particles with energies exceeding 100 GeV and would introduce as well a phase lag. However, the nature of void bubbles puts this matter in doubt. A void bubble is not a particle. Because it is instead a feature of the spacetime metric, it can be both massless *and* nonrelativistic. It is, moreover, a boundary on spacetime at which Dirichlet boundary conditions hold. Consequently, particles cannot penetrate the bubbles, but nor can they be reflected by it. Reflection would necessarily change the momentum of the particle, which would require momentum to be imparted to the void bubble. But this is not possible. The void bubble is a massless non-particle, at least in the ADM sense; it cannot receive momentum. After an encounter with a void bubble, a particle's momentum, then, would be unchanged. A shift in its phase could not be ruled out, however, because this does not violate the conservation of momentum or energy. This shift would, moreover, be expected to increase as a wavelength of the propagating particle decreases.

To be more precise, it is only impossible to transfer momentum to a void bubble, when it is caricatured as a spherically symmetric object in minisuperspace with a single degree of freedom. In a classical model with all degrees of freedom accessible, an incident particle would deform the matter distribution at the surface of the bubble. With spherical symmetry thus violated, the matter would be free to emit a burst of momentum-carrying gravitational waves as it returns to its original symmetrical distribution. This would allow the incident particle to be deflected by the bubble without incurring a violation of momentum conservation. Quantum mechanically, the matter at the surface of the bubble is a stiff 2-brane permitting only discrete modes of deformation. On interacting with an incident particle it would not deform, unless the particle's momentum were large enough to impart sufficient energy to reach at least the first of these discrete modes. The energy of this lowest deformed mode depends inversely on a power of the brane's tension. Hence, it is easy to imagine a brane being sufficiently stiff to be immune to deformation by particles incident at energies available in the current generation of accelerators. In this case, the minisuperspace description forbidding momentum transfer to the void bubble would be adequate.

In extending the model to the case of (ADM) massive void bubbles there is a trade off. We would hope to find a slight mass that would account for the missing matter of the universe without creating an angular dispersion of radiation beyond the limits established by extant data. As discussed above, the existence of such dispersion necessarily follows from the assumption of massive void bubbles. Demonstrations, then, of the absence of dispersions in either angle or phase of high-energy particle or radiation beams would be an experimental means through which the idea of slightly massive void bubbles pervading the whole of space could be refuted.

A spacetime densely packed with void bubbles would seem in essence to be a lattice theory. As such, it would be expected to violate local Lorentz invariance in that it introduces a special frame – the one in which the lattice is at rest. While it is true that the void bubbles introduce a special reference frame, so do the particles of matter in the universe. The zero-momentum frame for these particles is in some sense special, but its existence does not imply a problematic violation of Lorentz invariance. Similarly, void bubbles, like particles but unlike elements of a static lattice, are free to move relative to each other. Their motion may be likened to those of water molecules, permitting currents and eddies despite being densely packed. The existence of void bubbles, then, breaks Lorentz symmetry but in the same benign manner as does the existence of the ubiquitous cosmic microwave background and that of interstellar hydrogen atoms, or, more appropriately, as this symmetry is broken with regard to modes of vibration within a dense fluid or neutron star. In the latter cases one might have reason to expect the sort of Lorentz symmetry that characterizes acoustic manifolds [44]. The same cannot be said of explicit references in the model to manifestly non-Lorentz-invariant quantities such as the Planck volume. These references, however, may be restated as specifications of a ratio of volumes – the fraction of an arbitrary volume that consists of normal space to that consumed by void bubbles. While volumes themselves are not Lorentz invariant, their ratios are. Specification, moreover, of the size of the dominant void bubbles breaks Lorentz symmetry to the same degree as do specifications of the sizes of black holes or proton – not at all.

The value of this ratio – the fraction of normal space remaining in an apparent volume – is difficult to justify. Intuitively one suspects that proliferative bubble nucleation would not consume the whole of spacetime. Presumably, some of it would remain, even its dimensionality were to become fractal. But how much? Here we have supposed that one Planck volume remains per dominant bubble volume. Unfortunately, there is no compelling argument for this supposition. It would, moreover, require modification, if the model were to accommodate a hypothetical GUT-scale vacuum energy.

An advantage of the inelegance of this toy model, as opposed to optimistically ambitious integrations of Euclidean path integrals, is that the wormhole catastrophe is less severe and as a practical matter does not occur. The argument of Preskill [28], who extended Coleman's model beyond the case of dilute wormhole densities, suffices. Large wormholes (i.e. void bubbles) are unlikely to form, because the space they would occupy is almost certainly occupied by small wormholes. In the absence of infinite spikes in the probability density obtained through functional integration, Polchinski's counter [8] to Preskill does not apply. A finite suppression of the large wormhole probability is all that is required. Although an unattractive feature, the manual insertion of a length scale would seem, moreover, to grant immunity to the renormalization group scaling arguments [9] that bring the wormhole catastrophe to light. Lastly, if a large wormhole were somehow to form, superior pressure from its neighbors would force it to contract until its size matched that of wormholes with which it is contiguous.

Another advantage of this form of mass bubble nucleation is that the bubbles lower the value of the cosmological constant immediately. Unlike the usual nucleation scenarios, they need not slowly expand at the speed of light until they have engulfed a large fraction of the universe. They need only expand until they are no larger than one percent of the size of a proton.

We now have, moreover, an answer to a question posed many years ago by Roman, when he discovered that primordial wormholes inflate [40]: Where are they? The answer is that they are everywhere. By conforming to the brane world ethos and taking the wormholes to be brane boundaries on spacetime, i.e. void bubbles, we ensure that inflating wormholes no longer coalesce and do not thereby expand without limit. Their runaway expansion is instead halted by the collisions of the stiff branes that form their throats. This occurs when the void bubbles (semi-wormholes) are no larger than $10^{-15}$ cm.

Avenues for future work include: 1) explicitly demonstrating that higher-order terms in the gravitational Lagrangian can cause the wormhole potential to roughly resemble that in Figure 2, 2) determining the degree to which the cheap quantization employed here -- in which clarity and ease were purchased at the cost of rigor – approximates the formal methods that feature nonlocal differential operators, 3) taking the Casimir effect into account, 4) considering slightly massive void bubbles, 5) working out void-bubble-induced dispersion relations for high-energy particle or radiation beams, 6) using fractal modeling to determine the volume, remaining after full nucleation, of normal space relative to that of the dominant void bubbles, and 7) incorporating an explicit model of brane-brane interactions to describe the collisions between expanding void bubbles.

## 6. Conclusion

It might be possible to resolve the cosmological constant problem, if the actual volume of any given region of space is a tiny fraction of its apparent volume. This is tantamount to supposing that spacetime is in some sense effectively hollow. This supposition can be realized by assuming spacetime to be filled with massless, submicroscopic, brane-bounded, bubbles of nothing – void bubbles. These can be modeled as thin shelled semi-wormholes in minisuperspace, whose dynamics are determined by a potential function. At the Planck scale, where this classical potential function is believed to be incorrect, it is assumed to resemble a square well. This permits bubble nucleation to be treated analogously to alpha particle decay. Inflation appears a likely trigger for mass nucleation, whose denouement is a dense packing of dominant bubbles each of about $10^{-15}$ cm, the interstices of which contain smaller bubbles. The 2-brane at the boundary of each bubble prevents them from coalescing, imposes Dirichlet boundary conditions, and thereby ensures the predictability of physics. The nucleation process is assumed to continue until there remains a single Planck volume of normal space for each dominant void bubble. Continued bubble nucleation ensures that this ratio and the size of the dominant bubbles are maintained throughout the post-inflationary expansion of the universe. If the bubbles could be shown to be slightly massive, they would likely form the chief constituent of dark matter. It should be possible to work out dispersion relations through which the model could be subjected to experimental tests.

The key, then, to the puzzle of the cosmological constant might be a sort of hollow universe scenario in which empty space is emptier than we have heretofore believed.